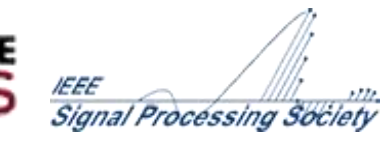

# NGSE-Corr: A technique for objective clinical evaluation of quantitative-imaging methods without a gold standard

Yan Liu, Ziping Liu, Zekun Li, Jingqin Luo, Daniel L. J. Thorek, Barry A. Siegel, Abhinav K. Jha, *Senior Member, IEEE*

***Abstract*—Objective evaluation of quantitative-imaging (QI) methods based on how reliably they measure true values is important for clinical translation. Performing such evaluation with patient data is highly desirable but hindered by the lack of gold standards. To address this challenge, advancing on previous studies, we propose a no-gold-standard evaluation technique, NGSE-Corr, that objectively evaluates QI methods without true values. The technique assumes a linear stochastic relationship between true and measured values, characterized by a slope, bias, and multivariate Gaussian-distributed noise term that models correlated noise across QI methods. We derive a maximum-likelihood approach to estimate these parameters using only measured values. From the estimates, we compute noise-to-slope ratio (NSR) to rank QI methods based on precision. Numerical experiments showed that NGSE-Corr reliably estimated the NSR, accurately ranked methods, and maintained performance even when assumptions made by the technique were partially violated. We also validated NGSE-Corr in an *in silico* imaging trial to rank three quantitative SPECT methods for measuring regional activity uptake in patients with bone metastatic castrate-resistant prostate cancer treated with radium-223. NGSE-Corr correctly identified the most precise QI method and ranked the methods for 95% (95% CI, 89%–98%) and 91% (95% CI, 84%–95%) of trials, respectively, with data from 50 patients. Performance further improved with larger cohorts. With 200 patients, NGSE-Corr yielded same rankings as those obtained with true values across all trial instances. These findings demonstrate the ability of NGSE-Corr to accurately rank QI methods without gold standards and motivate clinical validation and broader applications.**

## I. Introduction

QUANTITATIVE imaging (QI), i.e., the process of quantitatively extracting numerical features from medical images and using them to facilitate clinical decision making, is being actively investigated in many diagnostic and therapeutic procedures [1], [2]. These include estimating metabolic tumor volume from positron emission tomography (PET) to predict cancer therapy response [3], apparent diffusion coefficient from diffusion weighted magnetic resonance imaging for cancer staging [4], absorbed doses from single-photon emission computed tomography (SPECT) for dosimetry in radiopharmaceutical therapy (RPT) [5], mean density values from computed tomography images of lungs to monitor patients with COVID-19 [6] and radiomic features from medical images to develop biomarkers for cancer therapies [7]. Given the substantial clinical promise of QI, different QI methods have been and are being actively developed [8]–[11].

To realize the promise of QI, it is essential that measurements made through the QI methods are reliable, i.e., accurate and precise. A QI method that yields inaccurate measurements may not correctly reflect the underlying pathophysiology, while imprecise measurements may not provide confidence in making clinical decisions. Thus, objective evaluation of QI methods on their ability to yield reliable measurements is important. For clinical translation of these methods, such evaluation should be conducted ideally with patient data but is typically challenging due to the lack of gold standards. To circumvent this issue, physical-phantom and realistic simulation studies have been conducted [12]–[14]. These studies, while important, suffer from limitations that impact clinical confidence in their findings. For example, physical-phantom studies typically have limited ability to model physiology, anatomy, and patient-population variabilities, and realistic simulation studies may not

This work was supported by the National Institute of Biomedical Imaging and Bioengineering R01-EB031051 and R01-EB031962.

Yan Liu, Ziping Liu and Zekun Li are with the Department of Biomedical Engineering, Washington University, St. Louis, MO 63130 USA (e-mail: liu.yan@wustl.edu; liuziping@wustl.edu; zekunli@wustl.edu).

Jingqin Luo is with Division of Public Health Sciences, Department of Surgery, Division of Biostatistics and Alvin J. Siteman Cancer Center, Washington University, St. Louis, MO 63130 USA (email: jingqinluo@wustl.edu).

Daniel L. J. Thorek is with the Mallinckrodt Institute of Radiology, the Department of Biomedical Engineering, the Program in Quantitative Molecular Therapeutics and Alvin J. Siteman Cancer Center, Washington University, St. Louis, MO 63130 USA (e-mail: thorekd@wustl.edu).

Barry A. Siegel is with the Mallinckrodt Institute of Radiology and Alvin J. Siteman Cancer Center, Washington University, St. Louis, MO 63110 USA (e-mail: siegelb@wustl.edu).

Abhinav K. Jha is with the Department of Biomedical Engineering, the Mallinckrodt Institute of Radiology and Alvin J. Siteman Cancer Center, Washington University, St. Louis, MO 63130 USA (e-mail: a.jha@wustl.edu).

capture all biological and instrumentation factors [15]. Thus, there is an important need for techniques that can objectively evaluate QI methods with patient data.

Towards addressing this need, Hoppin et al [16], [17] proposed a regression-without-truth (RWT) technique for evaluating QI methods without gold standards. This technique is based on the premise that while the true quantitative values within the patient are unknown, the measured values result from imaging the patient and then processing the image data. The measured values are thus expected to be mathematically related to the true values. More specifically, the RWT technique posits a linear stochastic relationship characterized by slope, bias, and noise standard deviation terms, and estimates these terms using a maximum-likelihood-based approach without access to the true values. The RWT technique has since been further developed and evaluated in different applications[18], [19], [15], [20]. Overall, RWT provides a rigorous mathematical formulation to evaluate QI methods without gold standards.

For clinical translation, it is essential to advance this mathematical formulation to model clinically realistic scenarios. In this context, the existing RWT-based techniques assume that the noise between measurements of different QI methods are independent. This assumption is prone to violation in clinical settings because different QI methods typically process the same acquired projection or reconstructed image data, which contain measurement noise. As a result, the noise present in the measured data propagates through each QI method, leading to correlated noise in the resulting measurements. Thus, it is important to account for such correlation. To address this need, we advance upon this mathematical formulation to propose a no-gold-standard evaluation (NGSE) technique for objective assessment of QI methods without requiring true values when correlated noise is present. We term this technique as NGSE-Corr. In prior preliminary studies, we have explored this idea [21], [22] and here we place the idea on a firm mathematical footing with validations that comprehensively evaluate the technique across a range of numerical studies and in clinically realistic settings.

Clinical translation of this mathematical formulation also requires validation of the NGSE-Corr technique towards answering clinical questions. However, such validation would require access to true values, which are typically unavailable in clinical settings. Further, conducting such validations requires recruiting and scanning patients, which is time consuming, expensive, exposes patients to radiation risks, and has logistical challenges. The emerging paradigm of *in silico* imaging trials provides a mechanism to circumvent these challenges [13], [23], and identify whether NGSE-Corr would be a suitable candidate for clinical validation. Thus, in addition to numerical experiments, we conducted an *in silico* imaging trial to validate NGSE-Corr for evaluating quantitative SPECT (QSPECT) methods on the clinical task of measuring regional activity in patients with bone metastatic castrate-resistant prostate cancer treated with alpha-particle RPTs ($\alpha$-RPTs) based on $^{223}$Ra. This application is clinically important because toxicity to normal organs and tissues is a critical clinical concern in $\alpha$-RPTs. Thus, reliable quantification of activity uptake in radiosensitive organs is needed for assessing potential toxicities and guiding safe treatment planning [24]. Since the radionuclides also emit gamma rays that can be detected by a scintillation camera, QSPECT provides a mechanism to quantify the radiation-absorbed dose based on regional activity. However, quantification in $\alpha$-RPT SPECT is challenging due to the extremely low photon counts, complex emission spectra, and multiple image-degrading effects in SPECT, which lead to high noise and reduced precision in estimated regional activity uptake [25]. Consequently, multiple QSPECT methods are being actively developed for this challenging quantification task [26], [27]. Clinical translation of these methods requires objectively evaluating them on the task of regional uptake quantification. Our validation of NGSE-Corr for this important clinical question and on a challenging quantitative task in a clinically realistic setting by conducting an *in silico* imaging trial provides a stringent test bed to evaluate NGSE-Corr and provides strong impetus towards clinical translation of this technique.

## II. Methods

### A. NGSE-Corr: Theory

Consider a population of $P$ patients who have been scanned by a medical-imaging system. A set of $K$ QI methods are then used to measure a certain quantitative value of interest from these patient images. For the $p^{th}$ patient, denote the true value by $a_p$ and the measured value obtained with the $k^{th}$ method by $\hat{a}_{p,k}$. The goal of NGSE-Corr is to use only the measurements from the $K$ QI methods to objectively evaluate these on the task of measuring the true quantitative values. In this section, we present the theoretical formalism to achieve this goal.

For each QI method, we start with the premise that the true and measured values are stochastically related. From a methodological perspective, linearity of such a relationship is highly desirable. This is because the true quantitative value represents a certain biological phenomenon (e.g., tracer uptake within an organ) and a linear relationship between the true and measured values would ensure that a change in the true value corresponds to a linear change in the measured value. In fact, a test of linearity is often a requirement in assessing QI methods [2]. In multiple cases, the linearity also holds due to the imaging physics and quantitation process. As an example, imaging systems such as PET, SPECT, and fluorescence molecular tomography, are described by linear operators. Additionally, imaging modalities such as CT can also be modeled as approximately linear. Next, while some image reconstruction methods are performed by linear operators, even iterative methods, such as the penalized-likelihood tomography reconstruction methods, can be considered approximately linear for signal that is not affected by the nonnegativity constraint [28]. Finally, the process of quantification from the reconstructed image is linear for several features such as mean activity uptake. Prior work has also provided empirical support for such linear relationship. For example, in studies with I-131 SPECT, it was observed that a linear model was suitable to describe the relationship between true and mean regional activity uptake measured using OSEM-reconstruction-based quantification method that compensated for different image-degrading factors [15]. Based on these considerations, we model the relationship between measured and true values using a linear model characterized by a slope term and a bias term, denoted by $u_k$ and $v_k$, respectively, for the $k^{th}$ method.

Additionally, the random component of the error between the true and measured quantitative values, which arise in the imaging and quantification procedures, are quantified by a noise term, denoted by $\epsilon_{p,k}$ for the $p^{th}$ patient and the $k^{th}$ method. Since these procedures can be modeled as a sequence of random processes, from the central limit theorem, we consider the noise as a Gaussian-distributed random variable. Thus, for the $k^{th}$ QI method, we can write the relationship between $\hat{a}_{p,k}$ and $a_p$ as follows:

$$\hat{a}_{p,k} = u_k a_p + v_k + \epsilon_{p,k}, \tag{1}$$

where $\epsilon_{p,k}$ denotes a zero-mean Gaussian distributed random variable with standard deviation denoted by $\sigma_k$. Now we note that the noise between measurements yielded by different QI methods arises from measuring the same true values and should be correlated. We thus model the noise terms of different QI methods using a multivariate Gaussian distribution. Overall, we can write the relationship between the true value $a_p$ and measured values using the $K$ QI methods in vector notation as follows:

$$\begin{bmatrix} \hat{a}_{p,1} \\ \hat{a}_{p,2} \\ \vdots \\ \hat{a}_{p,K} \end{bmatrix} = \begin{bmatrix} u_1 & v_1 \\ u_2 & v_2 \\ \vdots & \vdots \\ u_K & v_K \end{bmatrix} \begin{bmatrix} a_p \\ 1 \end{bmatrix} + \begin{bmatrix} \epsilon_{p,1} \\ \epsilon_{p,2} \\ \vdots \\ \epsilon_{p,K} \end{bmatrix}, \tag{2}$$

where $\boldsymbol{\epsilon}_p = [\epsilon_{p,1}, \epsilon_{p,2}, \dots, \epsilon_{p,K}]^T$ denotes a random vector that is considered to be sampled from a zero-mean multivariate Gaussian distribution with a covariance matrix $\boldsymbol{\Sigma}$. The diagonal entries of $\boldsymbol{\Sigma}$, denoted by $\sigma_k^2$, represent the variance of the noise term for each method, and the off-diagonal entries, denoted by $\sigma_{k,k'}$, represent the covariance between the noise terms for methods $k$ and $k'$.

In Eq. (2), denote $[a_p, 1]^T$ by $\boldsymbol{A}_p$, $[\hat{a}_{p,1}, \hat{a}_{p,2}, \dots, \hat{a}_{p,K}]^T$ by $\widehat{\boldsymbol{A}}_p$, and the matrix consisting of the slope and bias terms by $\boldsymbol{\Theta}$. We can then write Eq. (1) compactly as follows:

$$\widehat{\boldsymbol{A}}_p = \boldsymbol{\Theta}\boldsymbol{A}_p + \boldsymbol{\epsilon}_p. \tag{3}$$

In Eq. (3), as mentioned earlier, $\epsilon_p$ is assumed to follow a zero-mean Gaussian distribution with covariance matrix $\boldsymbol{\Sigma}$. Thus, $\widehat{\boldsymbol{A}}_p$ follows a multivariate Gaussian distribution with mean given by $\boldsymbol{\Theta}\boldsymbol{A}_p$ and covariance matrix $\boldsymbol{\Sigma}$. Therefore, the probability of observing $\widehat{\boldsymbol{A}}_p$ given known $\{a_p, \boldsymbol{\Theta}, \boldsymbol{\Sigma}\}$, denoted by $\mathrm{pr}(\widehat{\boldsymbol{A}}_p | \{a_p, \boldsymbol{\Theta}, \boldsymbol{\Sigma}\})$, is given by

$$\mathrm{pr}(\widehat{\boldsymbol{A}}_p | \{a_p, \boldsymbol{\Theta}, \boldsymbol{\Sigma}\}) = \frac{1}{\sqrt{(2\pi)^K |\boldsymbol{\Sigma}|}} \exp\left\{-\frac{1}{2}(\widehat{\boldsymbol{A}}_p - \boldsymbol{\Theta}\boldsymbol{A}_p)^T \boldsymbol{\Sigma}^{-1} (\widehat{\boldsymbol{A}}_p - \boldsymbol{\Theta}\boldsymbol{A}_p)\right\}, \tag{4}$$

where $|\boldsymbol{\Sigma}|$ denotes the determinant of the covariance matrix $\boldsymbol{\Sigma}$. Calculating $\mathrm{pr}(\widehat{\boldsymbol{A}}_p | \{a_p, \boldsymbol{\Theta}, \boldsymbol{\Sigma}\})$ requires knowledge of the true value, $a_p$, which is unavailable in this problem. To circumvent this issue, the true values are considered to be sampled from a parametric probability distribution function (PDF) denoted by $\mathrm{pr}(a_p|\boldsymbol{\Omega})$. Multiplying both sides of Eq.(4) by $\mathrm{pr}(a_p|\boldsymbol{\Omega})$ and using the rule of conditional probability, we write the joint distribution of $\widehat{\boldsymbol{A}}_p$ and $a_p$ as

$$\mathrm{pr}(\widehat{\boldsymbol{A}}_p, a_p | \{\boldsymbol{\Theta}, \boldsymbol{\Omega}, \boldsymbol{\Sigma}\}) = \mathrm{pr}(\widehat{\boldsymbol{A}}_p | \{a_p, \boldsymbol{\Theta}, \boldsymbol{\Sigma}\}) \mathrm{pr}(a_p|\boldsymbol{\Omega}). \tag{5}$$

Marginalizing over $a_p$ on both sides of Eq. (5) yields

$$\mathrm{pr}(\widehat{\boldsymbol{A}}_p | \{\boldsymbol{\Theta}, \boldsymbol{\Sigma}, \boldsymbol{\Omega}\}) = \int \mathrm{pr}(\widehat{\boldsymbol{A}}_p | \{a_p, \boldsymbol{\Theta}, \boldsymbol{\Sigma}\}) \mathrm{pr}(a_p|\boldsymbol{\Omega}) da_p. \tag{6}$$

Next, denote $\widehat{\boldsymbol{\mathcal{A}}} = \{\widehat{\boldsymbol{A}}_p, p = 1, 2, \dots, P\}$ as the set of measurements yielded by the $K$ methods for all the $P$ patients. We assume that the $P$ true values are independent of each other. Then, we can write the distribution of $\widehat{\boldsymbol{\mathcal{A}}}$ as

$$\mathrm{pr}(\widehat{\boldsymbol{\mathcal{A}}} | \{\boldsymbol{\Theta}, \boldsymbol{\Sigma}, \boldsymbol{\Omega}\}) = \prod_{p=1}^{P} \int \mathrm{pr}(\widehat{\boldsymbol{A}}_p | \{a_p, \boldsymbol{\Theta}, \boldsymbol{\Sigma}\}) \mathrm{pr}(a_p|\boldsymbol{\Omega}) da_p, \tag{7}$$

Eq. (7) yields the likelihood of all the measurements in terms of only the linear-relationship parameters, $\{\boldsymbol{\Theta}, \boldsymbol{\Sigma}\}$, and the true-distribution parameters, $\boldsymbol{\Omega}$. Thus, calculating this likelihood does not require any knowledge of the true values. By maximizing the logarithm of this likelihood, we obtain the maximum-likelihood (ML) estimates of $\{\boldsymbol{\Theta}, \boldsymbol{\Sigma}, \boldsymbol{\Omega}\}$:

$$\{\widehat{\boldsymbol{\Theta}}, \widehat{\boldsymbol{\Sigma}}, \widehat{\boldsymbol{\Omega}}\}_{\mathrm{ML}} = \underset{\{\boldsymbol{\Theta}, \boldsymbol{\Sigma}, \boldsymbol{\Omega}\}}{\arg\max} \log\{\mathrm{pr}(\widehat{\boldsymbol{\mathcal{A}}} | \{\boldsymbol{\Theta}, \boldsymbol{\Sigma}, \boldsymbol{\Omega}\})\}$$
$$= \underset{\{\boldsymbol{\Theta}, \boldsymbol{\Sigma}, \boldsymbol{\Omega}\}}{\arg\max} \sum_{p=1}^{P} \log\left\{\int \mathrm{pr}(\widehat{\boldsymbol{A}}_p | \{a_p, \boldsymbol{\Theta}, \boldsymbol{\Sigma}\}) \mathrm{pr}(a_p|\boldsymbol{\Omega}) da_p\right\}. \tag{8}$$

The ML estimator has several properties that make it an optimal approach to estimate these parameters. Specifically, if an efficient estimator exists, the ML estimator is efficient, with variance approaching the Cramér-Rao lower bound. The schematic in Fig. 1 provides an intuition into the idea of the NGSE-Corr technique.

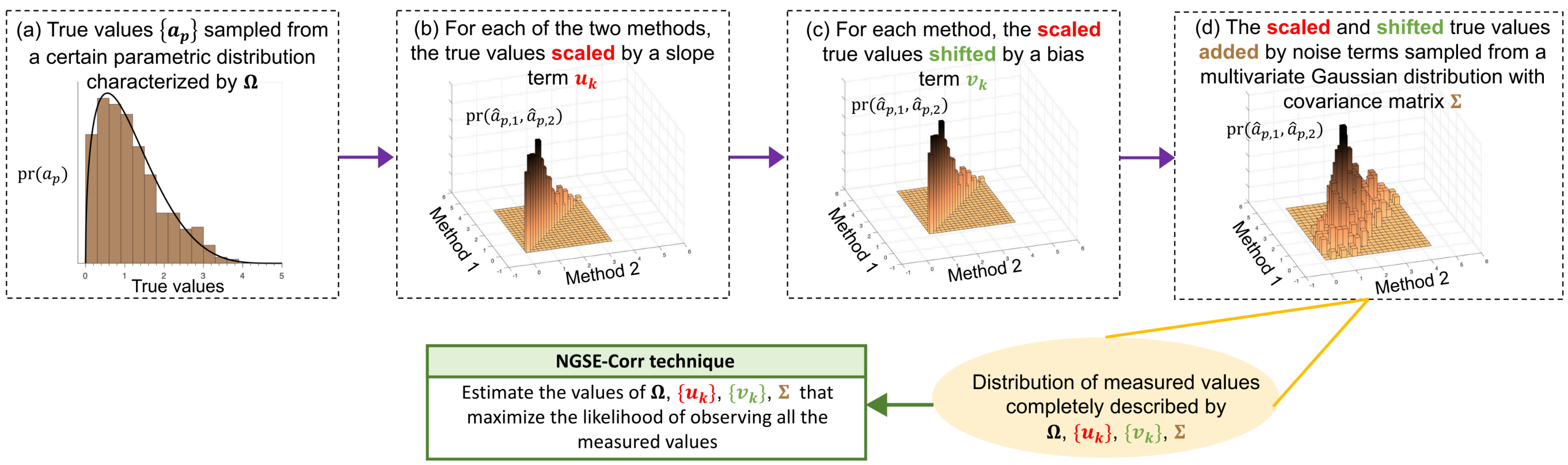

Fig. 1. Schematic illustrating the intuition behind the idea of the NGSE-Corr technique. We observe from (d) that the distribution of measured values obtained with the two methods can be completely described by the parameters that characterize the true-value distribution and the linear relationship between the true and measured values. Thus, theoretically, a maximum-likelihood-based approach can be designed to estimate these parameters from only the measurements and without any access to the true values.

Following estimation of these parameters, from Eq. (1), we note that the slope term, $u_k$, can be used to recalibrate the measured values, ensuring that the different QI methods are compared on a similar scale. This recalibration process would then lead to scaling of the noise standard deviation term $\sigma_k$, which quantifies precision of the $k^{th}$ QI method. Consequently, following the recalibration process, the ratio of the noise standard deviation term, $\sigma_k$, and the slope term, $u_k$, can be used to quantify the precision of the $k^{th}$ method. Thus, to rank the performance of the methods on the basis of precision, the noise-to-slope ratio (NSR), $\sigma_k/u_k$, is used as a figure of merit (FoM), with lower NSR values corresponding to higher precision. This choice of FoM is also consistent with prior RWT studies [16], [17].

### B. NGSE-Corr: Implementation

As described in Sec. II.A, we assume that the true quantitative values are sampled from a certain parametric PDF. This PDF should be able to model a wide variety of shapes and bounds of the distribution of true values. Further, true quantitative values are typically non-negative in medical imaging, and thus the chosen PDF should model this constraint. One PDF that satisfies these properties is the four-parameter beta distribution (FPBD). The FPBD is parameterized by $\mathbf{\Omega} = \{\alpha, \beta, g, l\}$, where $\{\alpha, \beta\}$ specify the shape, allowing for several types of unimodal distributions, including symmetric, non-symmetric, left or right-skewed, strictly increasing or decreasing, concave, convex and uniform distributions. Also, $\{g, l\}$ specify the upper and lower bound of the FPBD, respectively, defining the range of true values. The PDF of $a_p$, denoted by $\mathrm{pr}(a_p|\mathbf{\Omega})$, is given by

$$\mathrm{pr}(a_p|\mathbf{\Omega}) = \frac{(a_p - l)^{\alpha-1}(g - a_p)^{\beta-1}}{B(\alpha,\beta)(g-l)^{\alpha+\beta-1}}, \quad (9)$$

where $B(\alpha,\beta)$ denotes the beta function:

$$B(\alpha,\beta) = \int_0^1 x^{\alpha-1}(1-x)^{\beta-1}dx. \quad (10)$$

NGSE-Corr was implemented in MATLAB® (MathWorks, Natick, Mass). The ML estimates, $\{\hat{\mathbf{\Theta}}, \hat{\mathbf{\Sigma}}, \hat{\mathbf{\Omega}}\}_{\mathrm{ML}}$ in Eq. (8), were obtained using an iterative constrained optimization method based on the interior-point algorithm [29]. Our optimization routine ensured the covariance matrix, $\mathbf{\Sigma}$, to be positive definite by constraining the leading principal minors of $\mathbf{\Sigma}$ to be positive [30]. The search spaces for parameters $\{\mathbf{\Theta}, \mathbf{\Sigma}, \mathbf{\Omega}\}$ were set such that a wide range of shapes and bounds of the true distribution as well as expected relationships between the true and measured values could be modeled. The specific search-space values are provided in each experiment (Sec. III). In each iteration of the optimization routine, the values of $\{\mathbf{\Theta}, \mathbf{\Sigma}, \mathbf{\Omega}\}$ were initialized randomly to minimize the possibility of the algorithm being trapped in local minima. The iterations were stopped when either the changes in the estimated parameters between successive iterations fell below a specified threshold or the first-order optimality measure, specifically, the norm of the gradient of the likelihood function, was below a predefined tolerance. At this point, the estimated parameters were output by NGSE-Corr and used to derive the NSR to rank the QI methods.

## III. Validation of the NGSE-Corr technique

To validate NGSE-Corr, we conducted both numerical experiments and an *in silico* imaging trial. The numerical experiments validated NGSE-Corr under controlled setting where measurements of three hypothetical QI methods were generated with the assumptions of the technique (Sec. II) being satisfied. Different conditions were considered in numerical experiments, including different shapes and bounds of the true value distribution and varying levels of correlated noise among the QI methods. The *in silico* imaging trial, on the other hand, was designed to validate NGSE-Corr in a clinically realistic scenario using a digital phantom-based population and a simulated scanner, with three QI methods being evaluated. Finally, we also investigated the sensitivity of NGSE-Corr when assumptions made by this technique were violated.

### A. Validating NGSE-Corr with Numerical Experiments

In each numerical experiment, we sampled $P = 200$ true values from a known FPBD. From these true values, we generated synthetic measurements for $K = 3$ hypothetical QI methods, which were linearly related to the true values with correlated noise as described in Sec. II.A. These correlated noisy measurements were then input to NGSE-Corr to estimate $\{\mathbf{\Theta}, \mathbf{\Sigma}, \mathbf{\Omega}\}$, which were used to compute the NSR and rank the three QI methods based on precision.

#### *1) Performance for Different Shapes and Bounds of True Distribution*

To model different shapes of the distribution of true values, we considered three combinations of $\{\alpha, \beta\}$, namely, $\{1.5, 5\}$, $\{3, 3\}$, and $\{5, 1.5\}$. These combinations modeled left-skewed, symmetric, and right-skewed distributions, respectively. For each $\{\alpha, \beta\}$, we considered four combinations of upper and lower bounds, $\{g, l\}$, namely, $\{4.5, 0\}$, $\{4.67, 0.17\}$, $\{4.83, 0.33\}$, and $\{5, 0.5\}$.

From each set of true values, we generated noisy synthetic measured values for three hypothetical QI methods, where the values of slope, $\{u_k\}$, bias, $\{v_k\}$, and the noise variance, $\{\sigma_k^2\}$, were set to $\{1.10, 0.90, 1.05\}$, $\{0.10, 0.20, 0.30\}$, and $\{0.04, 0.09, 0.2025\}$, respectively. The covariances of noise terms between the different methods, $\{\sigma_{1,2}, \sigma_{1,3}, \sigma_{2,3}\}$, were set to $\{0.015, 0.02, 0.03\}$. These measurements were input to NGSE-Corr, which was tasked to estimate the NSR values for each method and then rank the QI methods based on precision.

To account for a wide range of shapes and bounds of true value distribution, we set the search spaces for $g$ and $l$ to $[4.5, 5]$ and $[0, 0.5]$, respectively. The search spaces for $\alpha$ and $\beta$ were both set to $[1, 20]$ to allow various shapes of true value distribution, including symmetric, non-symmetric, left or right-skewed distributions. To model different possible linear relationships between the true and measured values, we set the search spaces for $\{u_k\}$ and $\{v_k\}$ to [0.75, 1.25] and [0, 0.4], respectively. The range for slope parameter allows variation around unity, accommodating both underestimation and overestimation within a practically relevant interval. The search space for bias similarly allows a relatively wide range of possible values. Finally, the search spaces for diagonal and off-diagonal entries of covariance matrix $\mathbf{\Sigma}$ were set to[1e-4, 0.25] and [-0.02, 0.12], respectively, to capture a wide range of noise levels and covariance between noise terms.

These numerical experiments were repeated for 200 different noise realizations to compute the accuracy of NGSE-Corr, quantified as the percentage of realizations in which NGSE-Corr correctly ranks the methods and the percentage of realizations in which it correctly identifies the most precise method. The Wilson score confidence intervals were computed for these percentages and are reported as uncertainty measures. of the measured data to assess the robustness of NGSE-Corr.

#### 2) *Performance for Different Levels of Correlated Noise*

To model different levels of correlated noise between the three hypothetical QI methods, we considered nine combinations of the off-diagonal elements of the covariance matrix, $\{\sigma_{1,2}, \sigma_{1,3}, \sigma_{2,3}\}$. For the lowest level of correlated noise, we set $\{\sigma_{1,2}, \sigma_{1,3}, \sigma_{2,3}\}$ to $\{0.0040, 0.0090, 0.02025\}$. We then increased the correlated noise by scaling $\{\sigma_{1,2}, \sigma_{1,3}, \sigma_{2,3}\}$ using a factor that ranged from 1 to 5 with a step size of 0.5. In this experiment, the values of slope, bias, and the variance of the noise terms of the three QI methods were set the same as in Sec. III.A.1. The values of $\mathbf{\Omega}$ were set to $\{1.5,\ 5,\ 4.67, 0.17\}$. For each choice of $\{\sigma_{1,2}, \sigma_{1,3}, \sigma_{2,3}\}$, we repeated these experiments for 200 noise realizations. The search spaces for $\{\mathbf{\Theta}, \mathbf{\Sigma}, \mathbf{\Omega}\}$ were set the same as in Sec. III.A.1.

### B. *Validating NGSE-Corr with an In Silico Imaging Trial*

We considered the clinical scenario where patients with bone metastatic castrate-resistant prostate cancer (bmCRPC) were imaged using SPECT after being treated with $^{223}$Ra-based α-RPTs. For organ-based dosimetry, the goal of this imaging procedure is to measure the regional uptake of this isotope in radiosensitive organs to ensure that dose limits are not exceeded. Multiple QSPECT methods are developed for this purpose, and each method may yield different uptake estimates. In this setting, variability in these uptake estimates can directly impact the ability to make clinical decisions, making precision an important performance characteristic. Thus, the clinical question is to determine the method that yields the most precise uptake values. We conducted an *in silico* imaging trial titled ISIT-RIGHT (*in silico* imaging trial to evaluate regression without ground truth under correlated noise [ISIT-RIGHT]) to evaluate the efficacy of NGSE-Corr in answering this clinical question. The primary objective of ISIT-RIGHT was to evaluate the accuracy of NGSE-Corr in identifying the most precise QSPECT method for measuring the regional uptake values.

A large cohort of virtual patients were generated, and the regional activity uptake values for each patient were then measured using three QSPECT methods. This large cohort was used to conduct 100 instances of ISIT-RIGHT. For each trial instance, 50 patients (as determined by a power analysis described later) were drawn from this cohort and the correspon-ding measurements were input to NGSE-Corr. The most precise QSPECT method determined using NGSE-Corr was compared with that obtained when the true values were known across these 100 trials, and thus the accuracy of the NGSE-Corr in correctly identifying the most precise method was determined.

The secondary objectives of ISIT-RIGHT included evaluating the accuracy of NGSE-Corr in ranking the QSPECT methods and investigating the performance of NGSE-Corr for varying number of samples. The trial design is shown in Fig. 2, with elements of the trial described below.

#### 1) *Trial Population*

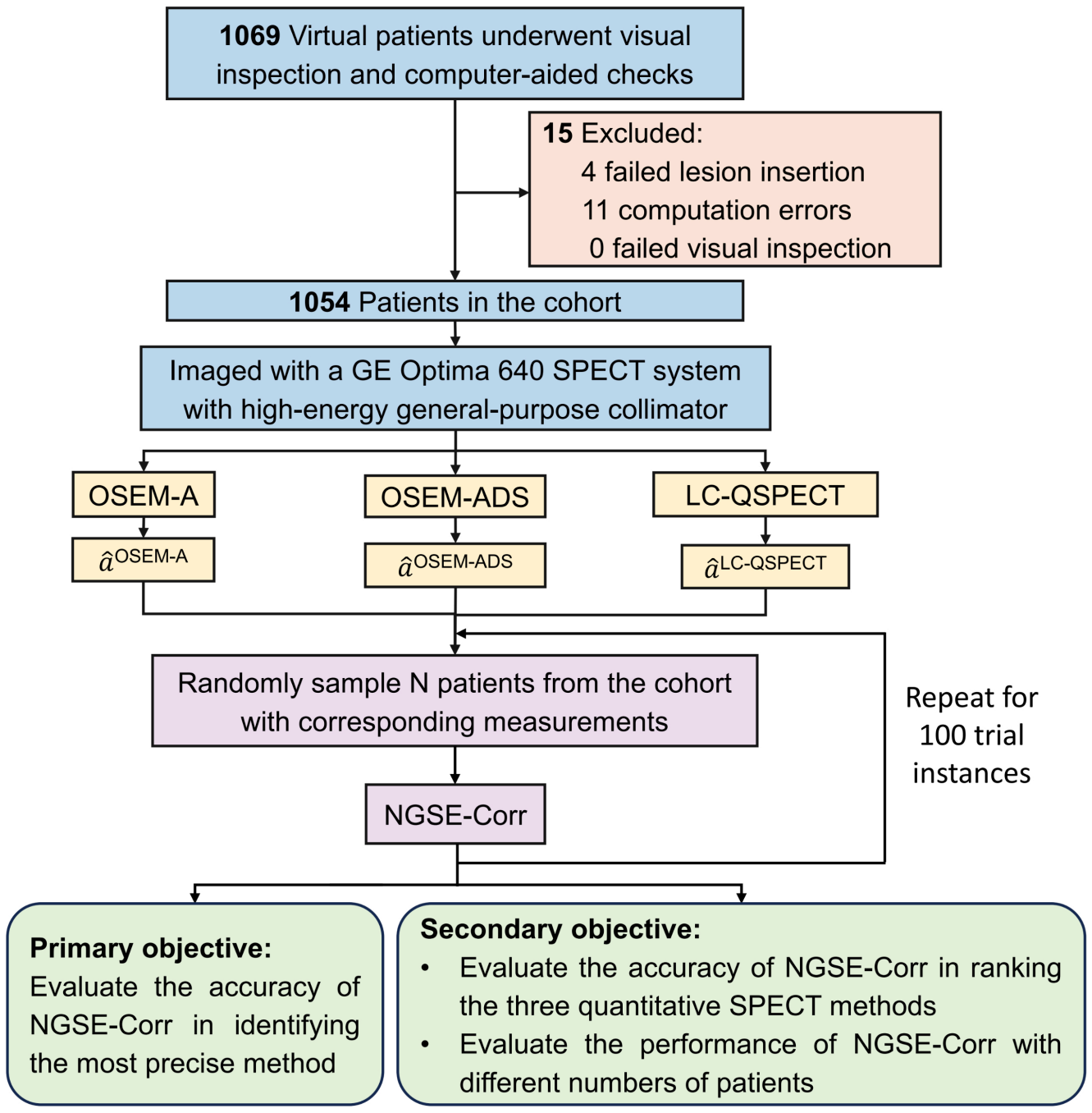

Fig. 2. Overall design of the ISIT-RIGHT.

The trial population was generated following the procedure described in Li et al [31]. We first simulated 1069 healthy virtual male patients using the 3D anthropomorphic Extended Cardiac-Torso (XCAT) phantom [32]. To model realistic anatomical variabilities in this population, we sampled the height and weight of each patient from the data published by the Centers for Disease Control and Prevention [33]. Lesions were inserted into each phantom with clinical-data derived characteristics, including lesion count, size, and spatial distribution [31]. Of these phantoms, 15 were removed due to computational errors, visual inspection and failure of lesion insertion. For each phantom, high-resolution 3D attenuation and activity maps (512 × 512 × 364 voxels, voxel size = 1.105 mm) were generated. From clinical SPECT images of patients treated with $^{223}$Ra-based $\alpha$-RPTs, we observe that organs with substantial $^{223}$Ra uptake include bone, small intestine, and large intestine, while the remaining organs have relatively low uptake, and we refer to them collectively as the background. Following the approach in [31], the uptake in background, bone, lesion, small and large intestines was independently sampled from Gaussian distributions, with a clinically derived mean uptake ratio of 2:5:20:25:25. Each patient received a single intravenous injection of $^{223}$Ra close to recommended dosage level [34]. The demographic and disease characteristics of simulated patients are shown in Table I.

#### 2) *Imaging Protocol*

The population was virtually imaged by a dual-head GE

Optima 640 SPECT system with a high-energy general-purpose collimator, using a well validated Monte Carlo simulation software, namely, SIMIND [35]. Photons were acquired from 60 uniformly spaced angular positions over 360°, in three energy windows: photopeak (68 – 102 keV) and two scatter windows (64 – 68 keV and 102 – 104 keV) [36]. At each angular position, the projection dimension was 128 × 128. The simulation modeled all relevant image-degrading processes in SPECT including photon attenuation and scatter, collimator response, septal penetration and scatter, characteristic X-rays, finite spatial and energy resolution of the detector, backscatter in the detector, and stray-radiation-related noise. For an average virtual patient, the SPECT projection had approximately 5,000 photons per axial slice, similar to clinical $^{223}$Ra-based $\alpha$-RPT SPECT acquisitions. The detailed SIMIND parameters used to simulate GE Optima 640 SPECT system are described in [31].

TABLE I
PATIENTS DEMOGRAPHICS AND CLINIC CHARACTERISTICS. CATEGORICAL CHARACTERISTICS ARE SUMMARIZED AS COUNTS AND PERCENTAGES; QUANTITATIVE CHARACTERISTICS ARE SUMMARIZED AS MEAN ± STANDARD DEVIATION.

| Characteristic | Data |
|---|---|
| **Number of patients** | 1054 |
| **Height (cm)** | 174.5 ± 7.8 |
| **Weight (Kg)** | 88.1 ± 20.4 |
| **BMI (Kg/m)** | |
| < 25 | 263 (25.0%) |
| 25 - 30 | 420 (39.8%) |
| ≥ 30 | 371 (35.2%) |
| **Mean uptake (kBq/mL)** | |
| Background | 0.09 ± 0.02 |
| Bone | 0.23 ± 0.06 |
| Small intestine | 1.08 ± 0.35 |
| Large intestine | 1.13 ± 0.40 |
| Lesion | 0.91 ± 0.28 |

### 3) *Quantitative SPECT Methods*

To measure the regional activity uptake, we considered three QSPECT methods. In each method, the true boundaries of the regions were assumed to be known. The first two methods quantify the regional uptake from reconstructed SPECT images. The reconstruction was performed using an ordered-subset expectation maximization (OSEM)-based approach implemented with the CASToR [37] software. The first method (OSEM-A) compensated for only attenuation (A), while the second method (OSEM-ADS) compensated for attenuation (A), collimator-detector response (D), and for scatter (S) using the triple-energy-window approach [38]. For both methods, the OSEM reconstruction was performed using 20 iterations and 6 subsets. The number of iterations was optimized by minimizing the normalized absolute error between the true and measured mean regional activity uptake values. These parameters are also consistent with those used in [31]. The reconstructed image dimensions were 128 × 128 × 91 (voxel side length = 4.418 mm). Finally, using the definition of the true boundaries, we computed the average uptake within each region on the reconstructed image to estimate the regional activity uptake.

The third method was a projection-domain quantitative SPECT method, namely low-count quantitative SPECT (LC-QSPECT) [25]. Unlike the reconstruction-based methods, the LC-QSPECT method directly quantifies the regional uptake from projection data. Details on the theory and implementation of the method are provided in Li et al [25].

### 4) *Power analysis*

To determine the required sample size (N in Fig. 2) for evaluating the accuracy of NGSE-Corr in identifying the most precise method, we performed a numerical power analysis. For this analysis, we first needed estimates of the linear relationship parameters of different QSPECT methods, which we obtained from a pilot realistic simulation study using 150 patients following a similar strategy outlined in Sec.III.(1)-(3), (5).

In the power analysis, we considered different sample sizes. For each sample size, synthetic measurements for three QSPECT methods were generated using the estimated linear relationship parameters and then input to NGSE-Corr to estimate the NSR for each method. As the pilot study indicated that LC-QSPECT had the lowest NSR, we computed the difference in NSR between each OSEM method and LC-QSPECT, i.e., $\text{NSR}_{\text{OSEM-A}}$-$\text{NSR}_{\text{LC-QSPECT}}$ and $\text{NSR}_{\text{OSEM-ADS}}$-$\text{NSR}_{\text{LC-QSPECT}}$ and calculated one-sided 95% confidence intervals (CIs) for these differences. This process was repeated 100 times per sample size. The proportion of trials in which the lower bounds of both CIs were greater than zero corresponds to the statistical power for NGSE-Corr to identify LC-QSPECT as the most precise method. We found that a sample size of 200, as could be obtained from N = 50 patients (four regions per patient) achieved 80% power to identify LC-QSPECT as the most precise method.

### 5) *Ranking QSPECT Methods Using NGSE-Corr*

We randomly sampled 50 patients without replacement from a cohort of 1,054 virtual patients. The regional uptake values from four regions, background, bone, and the small and large intestines, were combined, yielding 200 samples as input to the NGSE-Corr technique. The search spaces for $\{u_k\}$ was set to [0.75, 1.25]. The search space for $\{v_k\}$ was set to [-0.1, 0.4] kBq/ml. This allowed modelling for biases arising in due to factors such as partial volume effects[39] and non-negativity constraint in the reconstruction algorithm[40]. These effects are known to introduce bias in the reconstructed voxel values, which can propagate to the mean regional activity values being measured. For the diagonal and off-diagonal entries of the covariance matrix $\mathbf{\Sigma}$, we set the search spaces to, $[1\text{e-}5, 0.02]$ kBq$^2$/ml$^2$, and $[-1\text{e-}4,\ 1.5\text{e-}3]$ kBq$^2$/ml$^2$, respectively. The search spaces for both $\alpha$ and $\beta$ were set to $[1, 20]$, and those for $g$ and $l$ were set to $[2.30, 2.50]$ kBq/ml and $[0, 0.2]$ kBq/ml, respectively. From the estimated standard deviation and slope, we obtained the NSR values for the three QSPECT methods, which were used to rank these methods based on the precision of estimating the true mean regional uptake.

The ranking yielded by NGSE-Corr was compared with that obtained using true activity uptake values, which we referred to as the true rankings. Given knowledge of the true activity uptake values and the linearity assumption in Eq. (2), we estimated the linear-relationship parameters between the true and measured values for each QSPECT method using least-squares regression, from which NSR values were computed to define the true ranking.

To assess the accuracy of NGSE-Corr in identifying the most precise QSPECT method and in ranking the methods, this entire process of sampling patients, applying NGSE-Corr, and comparing to the true ranking was repeated 100 times, yielding 100 instances of the ISIT-RIGHT trial (Fig. 2).

#### *6) Impact of Sample Size on NGSE-Corr Performance*

We observed from Sec. II.A that applying NGSE-Corr to evaluate three QI methods requires estimating a total of 16 parameters, which may require obtaining data from many patients [19], [41]. Thus, we set one of our secondary objectives as studying the impact of the number of patients on the performance of NGSE-Corr. For this purpose, we varied the number of patients input to the technique from 25 to 200. For each sample size, we performed 100 trial instances to assess the accuracy of NGSE-Corr in ranking the considered QSPECT methods and identifying the most precise method.

### *C. Evaluating Sensitivity of NGSE-Corr to Model Mismatch*

#### *1) Correlation Between True Values*

The derivation of NGSE-Corr (Eq. 7) assumes that true values are independent of each other, but this may not hold, for example, when we obtain multiple quantitative values from the same patient (as in our *in silico* imaging trial). Thus, we evaluated the sensitivity of NGSE-Corr when this assumption is violated.

We first sampled a set of 200 true values independently from a FPBD, where the values of $\mathbf{\Omega}$ were set to $\{1.5, 5, 2.43, 0.04\}$. To simulate correlation between the true values, we grouped the 200 true values into sets of four (corresponding to four regions in each patient) and filtered each set using a Gaussian kernel. The magnitude of the correlation was characterized by the standard deviation of the Gaussian kernel [42]. To model different degrees of correlation, we varied the standard deviation from 0 to 1. From the correlated true values, we generated synthetic measurements for three hypothetical QI methods, where the values of slope, bias, diagonal and off-diagonal elements of the covariance matrix were set to $\{1.18, 0.82, 1.00\}$, $\{0.249, \text{-}0.42\text{e-}2, 0.3\text{e-}3\}$, $\{1.63\text{e-}2, 5.64\text{e-}3, 7.8\text{e-}5\}$ and $\{1.25\text{e-}3, 6.73\text{e-}5, 1.66\text{e-}5\}$, respectively. The values of these parameters were derived from ISIT-RIGHT to evaluate the sensitivity of NGSE-Corr in a clinically realistic setting. For each choice of the standard deviation of the Gaussian kernel, we repeated the experiment for 200 noise realizations. The search spaces for $\{\mathbf{\Theta}, \mathbf{\Sigma}, \mathbf{\Omega}\}$ were set the same as in Sec. III.B.5.

#### *2) Nonlinearity Between True and Measured Values*

Another assumption made by NGSE-Corr is that the true and measured values are linearly related. While this linearity is a desired trait, we studied the performance of the technique when such linearity is not satisfied.

We first sampled 200 true values from an FPBD, with the parameters $\mathbf{\Omega}$ set as in Sec. III.C.1. Similarly, three hypothetical QI methods were considered with slope parameters $\{u_k\}$, bias parameters $\{v_k\}$, and covariance matrix entries of $\mathbf{\Sigma}$ set as in Sec. III.C.1. For one of the QI methods, the relationship between the measurements and the true values was modeled as quadratic. To introduce different degrees of nonlinearity, we varied the second-order coefficient in the quadratic relationship, over the range -0.25 to 0.75. For each choice, NGSE-Corr was executed for 200 noise realizations. The search spaces for $\{\mathbf{\Theta}, \mathbf{\Sigma}, \mathbf{\Omega}\}$ were again set the same as in Sec. III.C.1.

#### *3) Multimodal Distribution of True Values*

In the implementation of NGSE-Corr, we considered an experimental setup where the true values are sampled from an FPBD. While the FPBD is effective at modeling unimodal distributions, the distribution of true values in clinical studies could be multimodal, for example, due to including patients at different stages of the disease. To investigate the sensitivity of NGSE-Corr when such multimodality is present, we perturbed the FPBD using a Gaussian mixture model (GMM). More specifically, we modeled the true value distribution as a mixture of the FPBD and GMM such that

$$\mathrm{pr}(a_p|\mathbf{\Omega}, \{\mu_m\}, \{\sigma_m\}) = (1-\omega)\frac{(a_p - l)^{\alpha-1}(g - a_p)^{\beta-1}}{B(\alpha,\beta)(g-l)^{\alpha+\beta-1}} + \frac{\omega}{M}\sum_{m=1}^{M}\frac{\exp\left[-\frac{1}{2}\left(\frac{a_p-\mu_m}{\sigma_m}\right)^2\right]}{\sigma_m\sqrt{2\pi}}, \quad (11)$$

where $B(\alpha,\beta)$ is given by Eq. (10), $\mu_m$ and $\sigma_m$ denote the mean and standard deviation of the $m^{th}$ component of the GMM, and $\omega$ denotes the weight of GMM. The values of $\mathbf{\Omega}$ were set the same as in Sec.III.C.1. The GMM consisted of $M = 2$ components, with means of $\{1.8, 2.0\}$ and standard deviations of $\{0.3, 0.3\}$. The value of $\omega$ was varied between 0 and 0.4 to model different extents of perturbation.

For each choice of $\omega$, we sampled 200 true values from the perturbed distribution. We generated synthetic measurements from these true values for three hypothetical QI methods, with the values of slope, bias, and elements of the covariance matrix set the same as in Sec.III.C.1. The experiment was repeated for 200 noise realizations for each $\omega$. The search spaces for $\{\mathbf{\Theta}, \mathbf{\Sigma}, \mathbf{\Omega}\}$ were again set the same as in Section III.C.1.

## IV. Results

### *A. Validating NGSE-Corr with Numerical Experiments*

#### *1) Different Shapes and Bounds of True Distribution*

Table II summarizes the mean and standard deviation of the estimated slope, noise standard deviation, and the resultant NSR of the three hypothetical QI methods for three different combinations of $\{\alpha, \beta\}$ (Sec. III.A.1). We observe that NGSE-Corr yields accurate estimates of the slope and noise standard deviation terms, and hence the NSR values. Additionally, the technique yielded the same ranking as that obtained with the knowledge of true quantitative values for 77% (95% CI, 74%-80%), 53% (95% CI, 50%–56%), and 81% (95% CI, 78%–84%) of the 800 experiments for each combination of $\{\alpha, \beta\}$ as listed in Table II, respectively. Further, the technique correctly identified the most precise method for 96% (95% CI, 94%–97%), 75% (95% CI, 72%–78%), and 94% (95% CI, 92%–95%) of the experiments, respectively.

TABLE II
THE MEAN AND STANDARD DEVIATION OF ESTIMATED SLOPES, NOISE STANDARD DEVIATIONS, AND THE RESULTANT NOISE-TO-SLOPE RATIOS (NSR) BY NGSE-CORR FOR DIFFERENT COMBINATIONS OF $\{\alpha, \beta\}$.
EST.: ESTIMATED; STD. DEV.: STANDARD DEVIATION.

| | Method 1 | Method 2 | Method 3 |
|---|---|---|---|
| True slope | 1.10 | 0.90 | 1.05 |
| True noise std. dev. | 0.20 | 0.30 | 0.45 |
| True NSR | 0.18 | 0.33 | 0.43 |

| $\{\alpha,\beta\}=\{1.5,5\}$ | | | |
|---|---|---|---|
| Est. slope | $1.12\pm0.10$ | $0.90\pm0.10$ | $1.06\pm0.11$ |
| Est noise std. dev. | $0.16\pm0.09$ | $0.30\pm0.06$ | $0.44\pm0.05$ |
| Est. NSR | $0.15\pm0.08$ | $0.34\pm0.09$ | $0.42\pm0.07$ |
| $\{\alpha,\beta\}=\{3,3\}$ | | | |
| Est. slope | $1.09\pm0.10$ | $0.88\pm0.09$ | $1.04\pm0.10$ |
| Est noise std. dev. | $0.19\pm0.11$ | $0.31\pm0.11$ | $0.43\pm0.09$ |
| Est. NSR | $0.18\pm0.12$ | $0.36\pm0.13$ | $0.42\pm0.10$ |
| $\{\alpha,\beta\}=\{5,1.5\}$ | | | |
| Est. slope | $1.14\pm0.07$ | $0.93\pm0.06$ | $1.08\pm0.06$ |
| Est noise std. dev. | $0.16\pm0.08$ | $0.29\pm0.08$ | $0.45\pm0.05$ |
| Est. NSR | $0.15\pm0.08$ | $0.32\pm0.09$ | $0.42\pm0.05$ |

### 2) *Performance for Different Levels of Correlated Noise*

Table III presents the mean and standard deviation of the estimated NSR for nine different levels of correlated noise (Sec. III.A.2). We observe that NGSE-Corr yielded accurate estimates of the NSR values for most noise levels. Additionally, the technique yielded the same ranking of the QI methods as that obtained when the true quantitative values were known for 85% (95% CI, 83%-87%) of the 1,800 experiments. Further, the technique correctly identified the most precise method for 94% (95% CI, 93%-95%) of the experiments.

TABLE III
THE MEAN AND STANDARD DEVIATION OF ESTIMATED NOISE-TO-SLOPE RATIOS (NSR) BY NGSE-CORR AT DIFFERENT CORRELATED NOISE LEVELS (1 = LOWEST, 9 = HIGHEST).

| | **Method 1** | **Method 2** | **Method 3** |
|---|---|---|---|
| True NSR | 0.18 | 0.33 | 0.43 |
| Estimated NSR (correlated noise level = 1) | $0.15\pm0.08$ | $0.35\pm0.08$ | $0.45\pm0.08$ |
| Estimated NSR (correlated noise level = 2) | $0.14\pm0.08$ | $0.33\pm0.07$ | $0.42\pm0.07$ |
| Estimated NSR (correlated noise level = 3) | $0.14\pm0.09$ | $0.35\pm0.06$ | $0.49\pm0.07$ |
| Estimated NSR (correlated noise level = 4) | $0.17\pm0.10$ | $0.35\pm0.08$ | $0.44\pm0.07$ |
| Estimated NSR (correlated noise level = 5) | $0.14\pm0.08$ | $0.31\pm0.07$ | $0.42\pm0.08$ |
| Estimated NSR (correlated noise level = 6) | $0.18\pm0.08$ | $0.34\pm0.07$ | $0.43\pm0.06$ |
| Estimated NSR (correlated noise level = 7) | $0.18\pm0.10$ | $0.35\pm0.09$ | $0.47\pm0.08$ |
| Estimated NSR (correlated noise level = 8) | $0.19\pm0.08$ | $0.36\pm0.08$ | $0.47\pm0.08$ |
| Estimated NSR (correlated noise level = 9) | $0.15\pm0.08$ | $0.30\pm0.07$ | $0.40\pm0.08$ |

## B. *Validating NGSE-Corr with an In Silico Imaging Trial*

### 1) *Ranking the Quantitative SPECT Methods*

Fig. 3 presents the estimated NSR values when the true regional uptake values are known and when NGSE-Corr is applied across 100 trial instances with 50 patients (Sec. III.B.5). We observe that NGSE-Corr estimated the NSR values accurately. Additionally, for 95% (95% CI, 89%-98%) of the 100 trial instances, the technique identified the same most precise method as that obtained using the true regional uptake values, and for 91% (95% CI, 84%–95%) of the trial instances, the technique yielded the same ranking.

### 2) *Impact of Sample Size on NGSE-Corr Performance*

Fig. 4 presents the performance of NGSE-Corr for different numbers of patients (Sec. III.B.6). As expected, the variance of the NSR estimated using NGSE-Corr decreased as the number of patients increased (Fig. 4 (a)). Notably, when 200 patients were considered, NGSE-Corr yielded the same ranking and identified the same most precise QSPECT method as when the true regional uptake values were used in all trial instances (100%, 95% CI, 96%–100%). The comparison of estimated NSR values in this case is shown in Fig. 5.

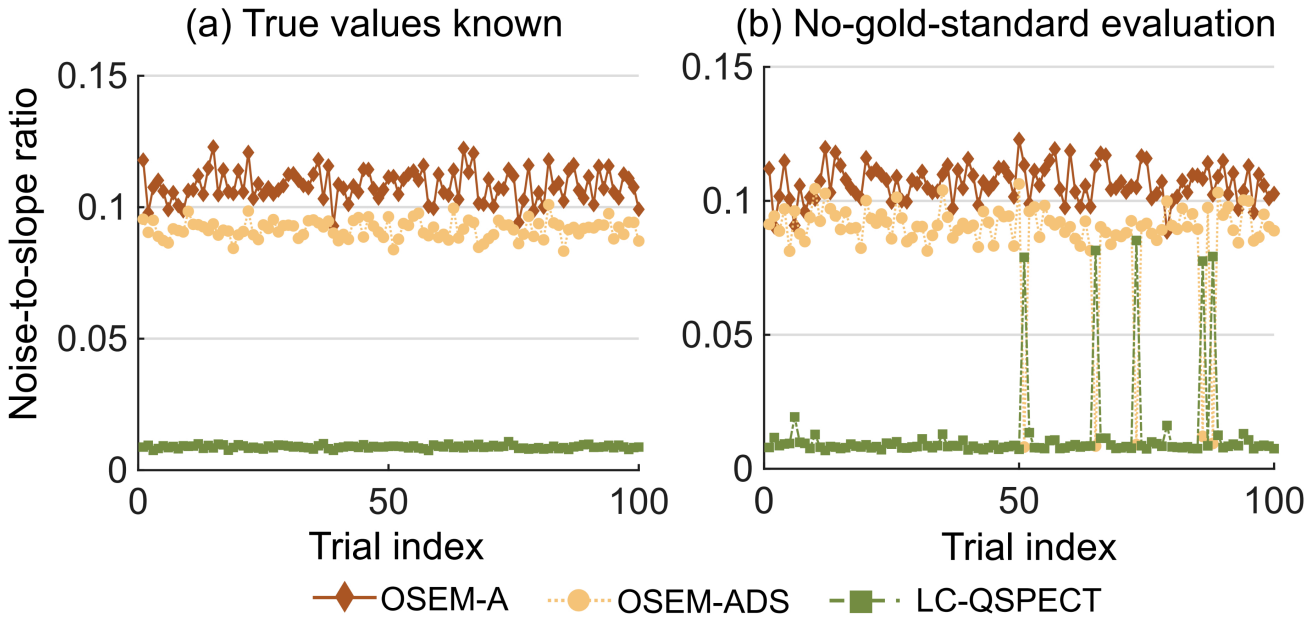

Fig. 3. The estimated NSR of the three considered QSPECT methods using 50 patients: (a) with true values known and (b) using the NGSE-Corr technique.

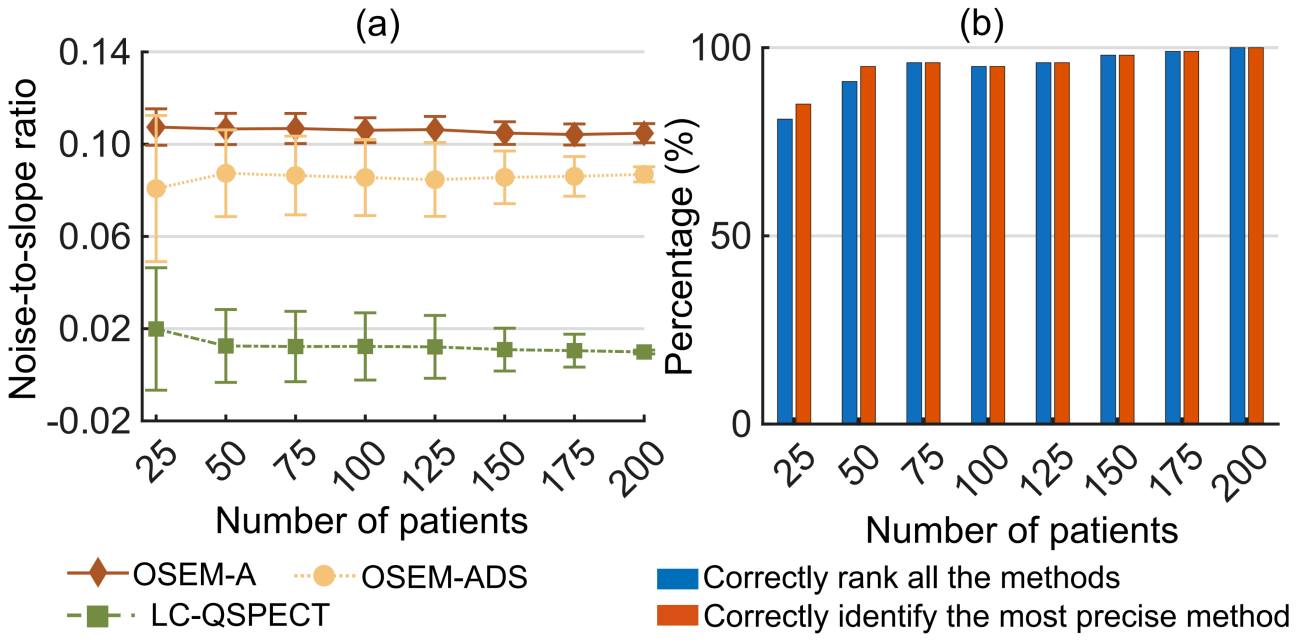

Fig.4. (a) The mean and standard deviation of the estimated NSR of the three considered QSPECT methods for different number of patient samples input to NGSE-Corr. (b) The percentage accuracy of NGSE-Corr in correctly ranking and identifying the most precise QI method.

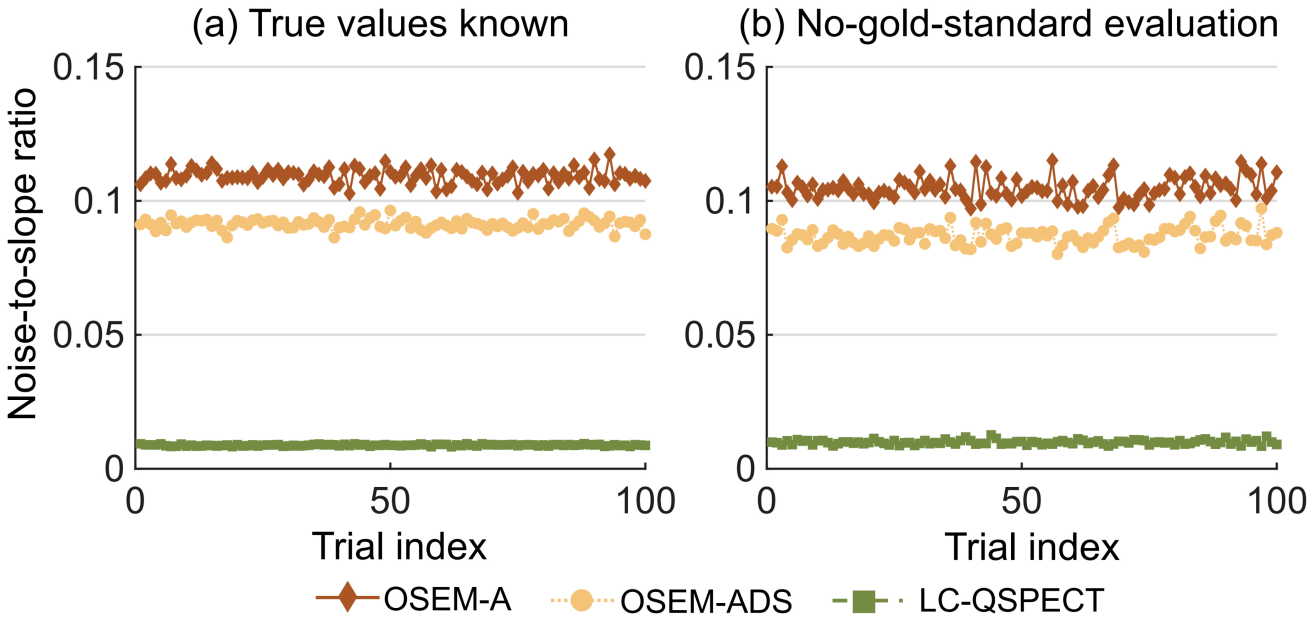

Fig. 5. The estimated NSR of the three considered QSPECT methods using 200 patients: (a) with true values known and (b) using the NGSE-Corr technique

## C. *Evaluating Sensitivity of NGSE-Corr to Model mismatch*

### 1) *Correlation Between True Values*

Fig. 6(a)-(b) show the performance of NGSE-Corr when different levels of correlation between the true values were introduced (Sec. III.C.3). It is observed that increases in the correlation did not lead to substantial variations of the performance in estimating the NSR and ranking the methods. Additionally, for each considered degree of correlation, NGSE-Corr correctly identified the most precise method in more than 95% of the noise realizations.

*2) Nonlinearity Between True and Measured Values*

Fig. 6(c)-(d) show the performance of NGSE-Corr when different degrees of nonlinearity between the true and measured quantitative values were considered (Sec. III.C.2). We observe that NGSE-Corr yielded accurate estimates of NSR for slight deviations from linearity. However, as the quadratic component of the relationship became more dominant, the estimates of NSR became less accurate. A similar trend was observed in ranking the methods and identifying the most precise method.

*3) Multimodal Distributions of True Values*

Fig. 6(e)-(f) present the performance of NGSE-Corr when different extents of mismatch between the actual and assumed distributions of the true values were considered (Sec. III.C.1). We observe that an increase in this model mismatch did not lead to substantial variations in the estimated NSR. Fig. 6(f) shows that for most considered weights of mismatch, NGSE-Corr correctly identified the most precise method for all the noise realizations and correctly ranked the three methods in more than 90% of the noise realizations.

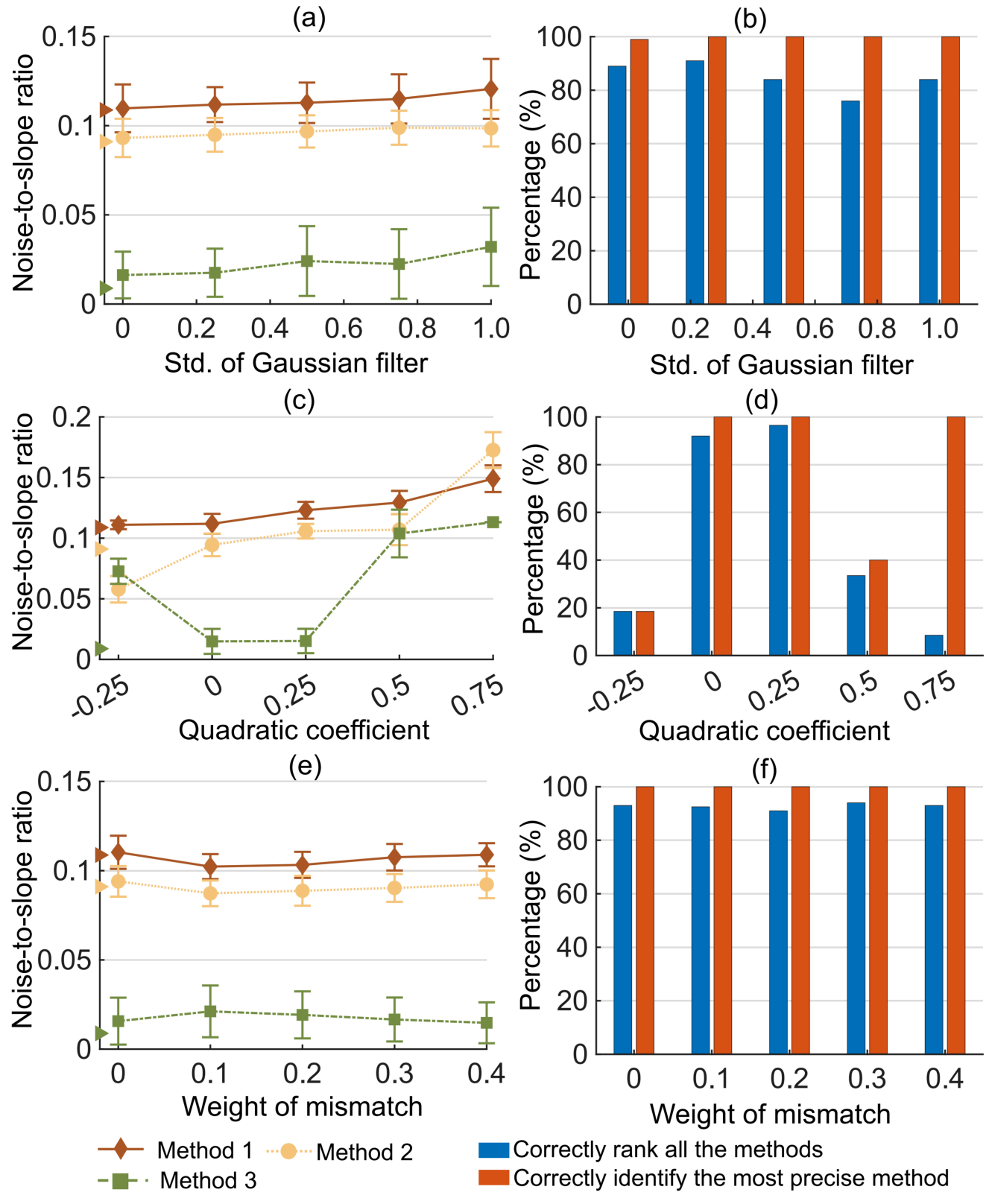

Fig. 6. The mean and standard deviation of the estimated NSR for (a) different extents of correlation between true values from the same patient; (c) different extents of non-linearity between the true and measured values; (e)different weights in the Gaussian mixture model to perturb the four-parameter beta distribution. The true NSR for each method is pointed out by arrows along the y-axis. Panels (b), (d) and (f) show the percentage accuracy in correctly ranking and correctly identifying the most precise QI method in corresponding cases.

## V. Discussion

The manuscript proposes and validates NGSE-Corr, a no-gold-standard evaluation technique for assessing QI methods on the basis of precision in the absence of ground truth. NGSE-Corr uses a theoretical formalism that accounts for correlated noise between measurements from different QI methods, thus modeling more clinically realistic settings and addressing a key limitation of previous RWT techniques. The results (Sec. IV) demonstrate the efficacy of this technique for evaluating QI methods across multiple scenarios.

Validation of NGSE-Corr using numerical experiments show that, when the assumptions made by this technique hold, the slope, noise standard deviation, and the resultant NSR values were estimated reliably across different shapes and bounds of the distribution of true values (Table II). Consequently, using NSR as the metric, the technique accurately ranked the QI methods and identified the most precise method. Further, the technique yielded consistently accurate ranking when different amounts of correlated noise were introduced between the QI methods (Table III). These results show that, when the assumption made by this technique holds, NGSE-Corr is effective in accurately ranking QI methods in the absence of ground truth.

An important aspect of our validation study was to evaluate NGSE-Corr beyond idealized numerical experiments to a clinically realistic scenario using an *in silico* imaging trial. This trial assessed the ability of NGSE-Corr to compare different QSPECT methods on the task of quantifying mean regional activity uptake. We observe from the results that, using data from 50 patients, NGSE-Corr accurately identified the most precise QSPECT method and ranked the considered QSPECT methods. Performance further improved as the number of patients was increased. Using data from 200 patients, NGSE-Corr correctly ranked the QI methods in all the trial instances (Fig. 4(b) and Fig. (5)). These results demonstrate the efficacy of NGSE-Corr for evaluating QI methods in a clinically realistic setting.

NGSE-Corr makes several assumptions that could potentially be violated in practice. Thus, we conducted studies to evaluate the sensitivity of NGSE-Corr to these assumptions. One such assumption is the independence of the true values. However, dependence may arise, for example, when multiple quantitative values are obtained from the same patient. Our results indicate that NGSE-Corr was relatively insensitive to dependencies between true values (Fig.6(a)-(b)). Another assumption is that true values follow a specific parametric distribution, which may again not hold in real settings. Our results show that NGSE-Corr was also relatively insensitive to the considered mismatches between the assumed and actual true value distributions (Fig. 6(e)-(f)). These observations are encouraging, as they support the use of multiple measurements from the same patient to help decrease the number of patient samples required for this technique.

The results also showed that while NGSE-Corr was relatively insensitive to slight deviations from linearity between the true and measured values, a high degree of nonlinearity could deteriorate the performance of the technique (Fig. 6(c)-(d)). This finding highlights the importance of assessing linearity between the true and measured values when applying NGSE-Corr. One approach is to assess this linearity through realistic simulations prior to applying the NGSE-Corr technique. Another option is to check for linearity between the measured values obtained with different QI methods. As shown previously, if the QI values obtained with the different methods are not linearly related, the true values will also not be linearly related to the measured value [20]. Finally, the NGSE-Corr technique assumes that the noise standard deviation is constant and independent of the magnitude of the true value. However, in medical imaging, measurement variance may depend on signal intensity, such that it cannot be represented by a single constant. Consequently, defining NSR as a single FoM may no longer be appropriate. Addressing this limitation represents an important direction for future work, including extending NGSE-Corr to account for signal-dependent noise and developing corresponding FoM.

In our *in silico* imaging trial, NGSE-Corr was applied to evaluate quantitative SPECT methods on the task of measuring mean regional activity uptake. However, this technique is broadly applicable and can also be used to evaluate QI methods in other applications, such as image denoising and segmentation. Often, in these applications, the QI methods are evaluated using metrics that may not necessarily correlate with performance on clinical tasks [13]. For example, in oncological PET, segmentation methods are typically evaluated using Dice scores, which quantify the spatial overlap between the true and predicted tumor segmentations. However, studies have shown that evaluating segmentation methods based on the Dice scores can lead to interpretations that are discordant with evaluation based on quantitative tasks [43]. Thus, there is an important need for objective task-based evaluation of these methods. NGSE-Corr provides a mechanism to conduct such evaluation with patient data without gold standards.

The performance of NGSE-Corr could be affected by the number of patient images input into the technique (Fig. 4). Therefore, guidance is needed to determine factors such as the required number of patients for the technique to yield accurate ranking. To address this need, we have developed a Cramér–Rao bound-based framework to quantify the upper bound on correctly ranking the QI methods without ground truth [44]. The framework can provide the best achievable ranking accuracy for a given set of QI methods with specific number of patients. If the upper bound is low, one approach is to collect more patient data for NGSE-Corr to yield higher accuracy. Further, to improve the performance of NGSE-Corr when even a small number of patients are available, one approach is to incorporate prior information about the parameters estimated by NGSE-Corr. A Bayesian-NGSE approach has shown promise to this end [45], and further development of such approaches are needed.

Next, we note that NGSE-Corr is developed to evaluate precision and not accuracy of QI methods. Accordingly, the use of NGSE-Corr is most appropriate in settings where precision is the performance measure relevant to the clinical task. Also, NGSE-Corr is a statistical technique and is not guaranteed to rank the QI methods correctly, as also seen in our results (Fig. 3). To address this issue, consistency checks can be performed to assess whether the parameters estimated using NGSE-Corr are consistent with the measured data [20], [41]. It should be noted that passing the checks does not guarantee that the technique is accurate. However, a failure of such check indicates that the output of NGSE-Corr needs to be interpreted with caution.

Finally, NGSE-Corr was validated with an *in silico* imaging trial and not with patient data. While our simulations were designed to be accurate and clinically realistic, they may not capture all aspects of system instrumentation and biology. Thus, further validation of NGSE-Corr with patient data is another important future direction.

## VI. Conclusion

We proposed a no-gold-standard evaluation technique for objective evaluation of quantitative-imaging (QI) methods with patient data without the knowledge of true quantitative values. The technique, referred to as NGSE-Corr, models correlated noise between the different QI methods Validation using numerical experiments demonstrates that NGSE-Corr yielded accurate rankings of the QI methods on the basis of precisely measuring the true values for a wide range of configurations. Results from an *in silico* imaging trial conducted in the context of evaluating quantitative SPECT methods for $^{223}$Ra-based radiopharmaceutical therapies (ISIT-RIGHT) showed the high accuracy of NGSE-Corr in a clinically realistic scenario. Further, NGSE-Corr yielded accurate ranking performance under moderate violations of several underlying assumptions, although larger deviations may impact performance. Overall, this study demonstrates the ability of NGSE-Corr to accurately evaluate QI methods even without ground truth. Our results motivate further validation with patient data and expanding the technique for a wider range of clinical applications.

Software to conduct this study is available at https://drive.google.com/drive/folders/1thL48YS4r97I6S9EjFirKAgG3NwA6dSf?usp=share_link.